\documentclass[aps,prb,a4paper,10pt,twocolumn,showpacs,floatfix,superscriptaddress,preprintnumbers,longbibliography]{revtex4-2}
\usepackage{float}
\usepackage{dcolumn,graphicx,color,booktabs,microtype,afterpage}
\usepackage{amssymb}
\usepackage{amsmath}
\usepackage[charter,greekuppercase=italicized]{mathdesign}
\usepackage{sidecap}
\usepackage[mathlines]{lineno}

\graphicspath{{./}{figure/}}
\renewcommand{\tablename}{Table}
\makeatletter\renewcommand{\fnum@figure}[1]{\figurename~\thefigure.~}\makeatother
\makeatletter\renewcommand{\fnum@table}[1]{\tablename~\thetable.}\makeatother

\newcount\hh \newcount\mm
\hh=\time \divide\hh by 60
\mm=\hh \multiply\mm by 60 \mm=-\mm
\advance\mm by \time
\def\now{\number\hh:\ifnum\mm<10{}0\fi\number\mm}

\usepackage[colorlinks,plainpages=false,linkcolor=blue,urlcolor=blue,citecolor=blue,pdfpagemode=UseNone,pdfstartview=FitBH]{hyperref}
\usepackage{nicefrac}

\newcommand{\tcr}[1]{\textcolor{black}{#1}}
\newcommand{\tcb}[1]{\textcolor{blue}{#1}}

\begin{document}

\makeatletter\renewcommand{\ps@plain}{%
\def\@evenhead{\hfill\itshape\rightmark}%
\def\@oddhead{\itshape\leftmark\hfill}%
\renewcommand{\@evenfoot}{\hfill\small{--~\thepage~--}\hfill}%
\renewcommand{\@oddfoot}{\hfill\small{--~\thepage~--}\hfill}%
}\makeatother\pagestyle{plain}

\preprint{\textit{Preprint: \today, \now}} 

\title{Fully-gapped superconductivity with preserved time-reversal symmetry in \\ NiBi$_3$ single crystals}
%
%
\author{T.\ Shang}\email[Corresponding authors:\\]{tshang@phy.ecnu.edu.cn}
\affiliation{Key Laboratory of Polar Materials and Devices (MOE), School of Physics and Electronic Science, East China Normal University, Shanghai 200241, China}
\affiliation{Chongqing Key Laboratory of Precision Optics, Chongqing Institute of East China Normal University, Chongqing 401120, China}

\author{J.\ Meng}
\affiliation{Key Laboratory of Polar Materials and Devices (MOE), School of Physics and Electronic Science, East China Normal University, Shanghai 200241, China}
%
%
%
%
%
%
\author{X.\ Y.\ Zhu}
\affiliation{Key Laboratory of Polar Materials and Devices (MOE), School of Physics and Electronic Science, East China Normal University, Shanghai 200241, China}
\author{H.\ Zhang}
\affiliation{Key Laboratory of Polar Materials and Devices (MOE), School of Physics and Electronic Science, East China Normal University, Shanghai 200241, China}
\author{B.\ C.\ Yu}
\affiliation{Key Laboratory of Polar Materials and Devices (MOE), School of Physics and Electronic Science, East China Normal University, Shanghai 200241, China}
\author{Z.\ X.\ Zhen}
\affiliation{Key Laboratory of Polar Materials and Devices (MOE), School of Physics and Electronic Science, East China Normal University, Shanghai 200241, China}
\author{Y.\ H.\ Wang}
\affiliation{Key Laboratory of Polar Materials and Devices (MOE), School of Physics and Electronic Science, East China Normal University, Shanghai 200241, China}
%
%
%
%
%
\author{Y.\ Xu}
\affiliation{Key Laboratory of Polar Materials and Devices (MOE), School of Physics and Electronic Science, East China Normal University, Shanghai 200241, China}
\author{Q.\ F.\ Zhan}
\affiliation{Key Laboratory of Polar Materials and Devices (MOE), School of Physics and Electronic Science, East China Normal University, Shanghai 200241, China}
\author{D.\ J.\ Gawryluk}
\affiliation{Laboratory for Multiscale Materials Experiments, Paul Scherrer Institut, CH-5232 Villigen PSI, Switzerland}
%
%
%
\author{T.\ Shiroka}
\affiliation{Laboratory for Muon-Spin Spectroscopy, Paul Scherrer Institut, CH-5232 Villigen PSI, Switzerland}
\affiliation{Laboratorium f\"ur Festk\"orperphysik, ETH Z\"urich, CH-8093 Z\"urich, Switzerland}

\begin{abstract}
We report a study of NiBi$_3$ single crystals by means of electrical-resistivity-, magnetization-, and muon-spin rotation and relaxation
($\mu$SR) measurements. As a single crystal, NiBi$_3$ adopts a needle-like
shape and exhibits bulk superconductivity with $T_c \approx 4.1$\,K. 
By applying magnetic fields parallel and perpendicular to the $b$-axis
of NiBi$_3$, we establish that its lower- and upper critical fields, as well as the magnetic penetration depths 
show slightly different values, suggesting a 
weakly anisotropic superconductivity. 
In both cases, 
the zero-temperature upper critical fields 
are much smaller than the Pauli-limit value, indicating that
the superconducting state is constrained by the orbital pair breaking.
The temperature evolution of the superfluid density, obtained from
transverse-field $\mu$SR, reveals
a fully-gapped superconductivity in NiBi$_3$, 
with a shared superconducting gap $\Delta_0$ = 2.1\,$k_\mathrm{B}$$T_c$ and magnetic penetration depths $\lambda_0$ = 223 and 210\,nm for $H \parallel b$-  and $H \perp b$, 
respectively. 
The lack of spontaneous fields below $T_c$ indicates that 
time-reversal symmetry is preserved in NiBi$_3$. 
The absence of a fast muon-spin relaxation and/or 
precession in the zero-field $\mu$SR spectra definitely
rules out any type of magnetic ordering in NiBi$_3$ single crystals.
Overall, our investigation suggests that NiBi$_3$ behaves as a conventional $s$-type superconductor. 
\end{abstract}

\maketitle\enlargethispage{3pt}

\vspace{-5pt}
\section{\label{sec:Introduction}Introduction}\enlargethispage{8pt}
Topological superconductors offer many attractive properties, which 
range from the possibility to host Majorana quasiparticles to enabling topological quantum computing~\cite{Qi2011,Kitaev2001,Chetan2008}. This has spurred researchers to 
explore different routes and/or materials 
to realize them. For instance, superconductors with $p+ip$ pairing have been predicted to support Majorana bound states at their vortices~\cite{Fu2008}. 
Besides relying on bulk superconductors with nontrivial electronic bands, another approach towards
topological superconductivity (SC) consists in combining conventional $s$-wave superconductors with topological insulators or ferromagnets, so as to form heterostructures. In both cases, proximity effects at the interface may lead to a two-dimensional superconducting state with an unconventional pairing~\cite{Peng2014,Bergeret2005,Buzdin2005}.

Among the different types of heterostructures, epitaxial Bi/Ni bilayers have been proposed as a promising candidate for hosting chiral $p$-wave SC~\cite{Wang2017a}.  
In fact, tunneling experiments in such
bilayers have shown the coexistence of ferromagnetism (FM) and SC below the superconducting transition, at $T_c \approx 4$\,K~\cite{Leclair2005}. 
More recently, surface magneto-optic Kerr-effect measurements revealed spontaneous magnetic fields  occurring below the onset of SC, 
which suggest the breaking of time-reversal symmetry (TRS) in the superconducting state of Bi/Ni bilayers~\cite{Gong2017}. Finally, time-domain terahertz spectroscopy experiments
found a nodeless SC state in Bi/Ni bilayers~\cite{Chauhan2019}, more consistent with a chiral $p$-wave SC~\cite{Chao2019}. 

At the same time, there is mounting evidence 
that the occurrence of SC in Bi/Ni bilayers could be
related to the formation of a superconducting NiBi$_3$ phase (with $T_c \approx$ 4\,K~\cite{Fujimori2000,Kumar2011,Silva2013}) during the thin-film growth~\cite{Siva2015,Siva2016,Wang2018,Bhatia2018,Bhatia2018,Liu2018,Vaughan2020,Das2022,Doria2022}. This because NiBi$_3$-free Bi/Ni bilayers 
show no signs of superconductivity~\cite{Liu2018,Vaughan2020}.  
These observations have triggered the researchers' interests to study the superconducting properties of pure NiBi$_3$.  
Similar to the Bi/Ni bilayers~\cite{Leclair2005}, also bulk NiBi$_3$ shows the coexistence of SC and FM~\cite{Liu2020,Pineiro2011,Herrmannsdorfer2011,Zhu2012}.
Thus, magnetization data in its superconducting state, show a clear loop of ferromagnetic hysteresis~\cite{Pineiro2011,Herrmannsdorfer2011,Silva2013}. 
Generally, superconductivity and ferromagnetism are antagonistic ground states. However, akin to certain U-based materials~\cite{Saxena2000,Aoki2001,Huy2007}, 
 NiBi$_3$, too, belongs to the rare class of ferromagnetic superconductors, potentially able to host spin-triplet pairing. 

To date, the superconducting properties of NiBi$_3$ have been mostly investigated via magnetic- and transport measurements~\cite{Fujimori2000,Kumar2011,Silva2013,Liu2020,Pineiro2011,Herrmannsdorfer2011,Zhu2012}. Yet, the microscopic nature of its SC, in particular, the superconducting order parameter, 
has not been explored and awaits  further investigation. In addition, it is not yet clear if the observed breaking of TRS and the unconventional SC of epitaxial Bi/Ni bilayers can be attributed to the formation of the NiBi$_3$ phase or not~\cite{Gong2017}. 

To clarify the above issues, we synthesized 
NiBi$_3$ single crystals using the flux method, and 
studied their superconducting properties by means of  electrical-resistivity-, magnetization-, and muon-spin rotation and relaxation ($\mu$SR) measurements.
We found that NiBi$_3$ exhibits a fully-gapped superconducting state with a preserved TRS. 
The absence of any magnetic order, on the surface or in the bulk of NiBi$_3$ single crystals, was also confirmed.  
Our results suggest that NiBi$_3$ behaves as a conventional $s$-type superconductor and, thus, the unconventional superconducting properties observed in Bi/Ni bilayers 
cannot be attributed to the NiBi$_3$ phase. 

\begin{figure}[!thp]
	\centering
	\includegraphics[width=0.46\textwidth,angle=0]{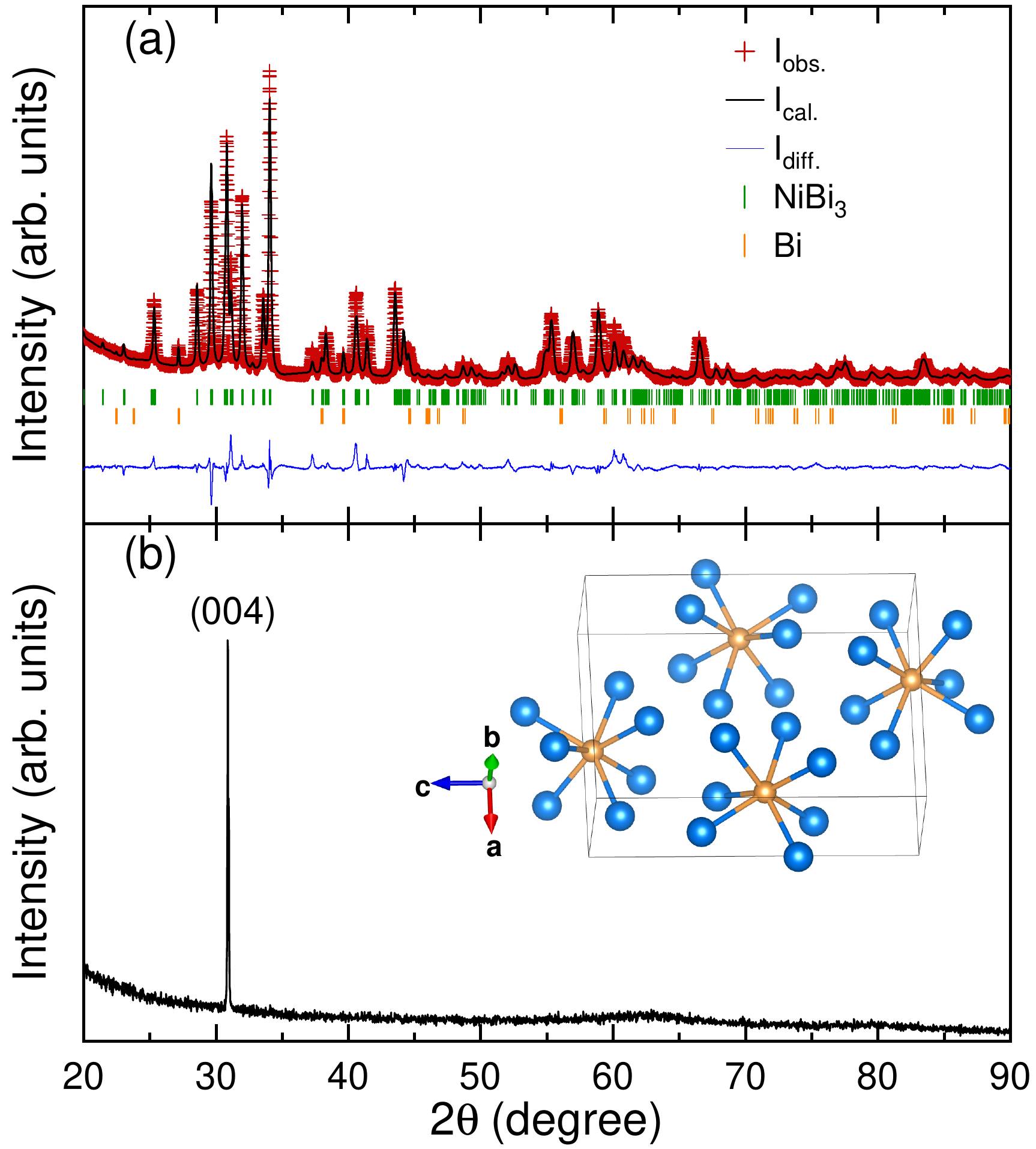}
	\caption{\label{fig:XRD} 
		(a) Room-temperature x-ray powder diffraction pattern and Rietveld refinement for NiBi$_3$. The red crosses and the solid
		black line represent the experimental pattern and the Rietveld refinement profile, respectively. The blue line at the bottom shows
		the residuals, i.e., the difference between the calculated and experimental data. The vertical bars mark the calculated Bragg-peak
		positions for NiBi$_3$ (green) and Bi (orange). The refinement $R$-factors
		are $R_\mathrm{p}$ $\sim$ 3.71~\%, $R_\mathrm{wp}$ $\sim$ 5.82~\%, and $R_\mathrm{exp}$ $\sim$ 1.37~\%. 
		Note that, the minority Bi phase comes from the remaining flux on the surface of crystals. (b) XRD pattern of a NiBi$_3$ single crystal.
		The unit cell is shown in the inset, with the blue
		and orange spheres depicting the Bi and Ni atoms.} 
\end{figure}
%

\section{Experimental details\label{sec:details}}\enlargethispage{8pt}
The NiBi$_3$ single crystals were grown from molten bismuth flux. 
High-purity Bi pieces (Alfa Aesar, 99.99\%) and Ni powders (Alfa Aesar, 99.9\%) with a 10:1 ratio were loaded in an Al$_2$O$_3$ crucible, which was sealed in a quartz ampoule under high vacuum. 
The quartz ampoule was heated up to 1050\,$^\circ$C and then kept at this temperature for over 24 hours. Finally, it 
was slowly cooled down to 360\,$^\circ$C at a rate of 2\,$^\circ$C/h, with the remaining Bi flux being removed via centrifugation.
The typical dimensions of the obtained needle-like single crystals were about  $\sim$3\,mm $\times$ 0.3\,mm $\times$ 0.3\,mm (see below). 
The crystals were checked by powder x-ray diffraction (XRD), recorded using a Bruker D8 diffractometer. 
Further characterization involved electrical-resistivity- 
and magnetization measurements, 
performed on a Quantum Design physical property measurement 
system (PPMS) and a magnetic property measurement system (MPMS), respectively. 
For the electrical-resistivity measurements, the dc current was applied along the $b$-axis.

The bulk $\mu$SR measurements were carried out at the multipurpose 
surface-muon spectrometer (Dolly) on the $\pi$E1 beamline of the Swiss muon source at Paul 
Scherrer Institut, Villigen, Switzerland. The crystals were aligned and mounted on 
a 25-$\mu$m thick copper foil, which ensured thermalization at low temperatures. The magnetic fields were applied both parallel and perpendicular to the $b$-axis of the NiBi$_3$ single crystal. 
The time-differential $\mu$SR data were collected upon heating and 
then analyzed by means of the \texttt{musrfit} software package~\cite{Suter2012}. 

%
\begin{figure}[!thp]
	\centering
	\vspace{-1ex}%
	\includegraphics[width=0.47\textwidth,angle=0]{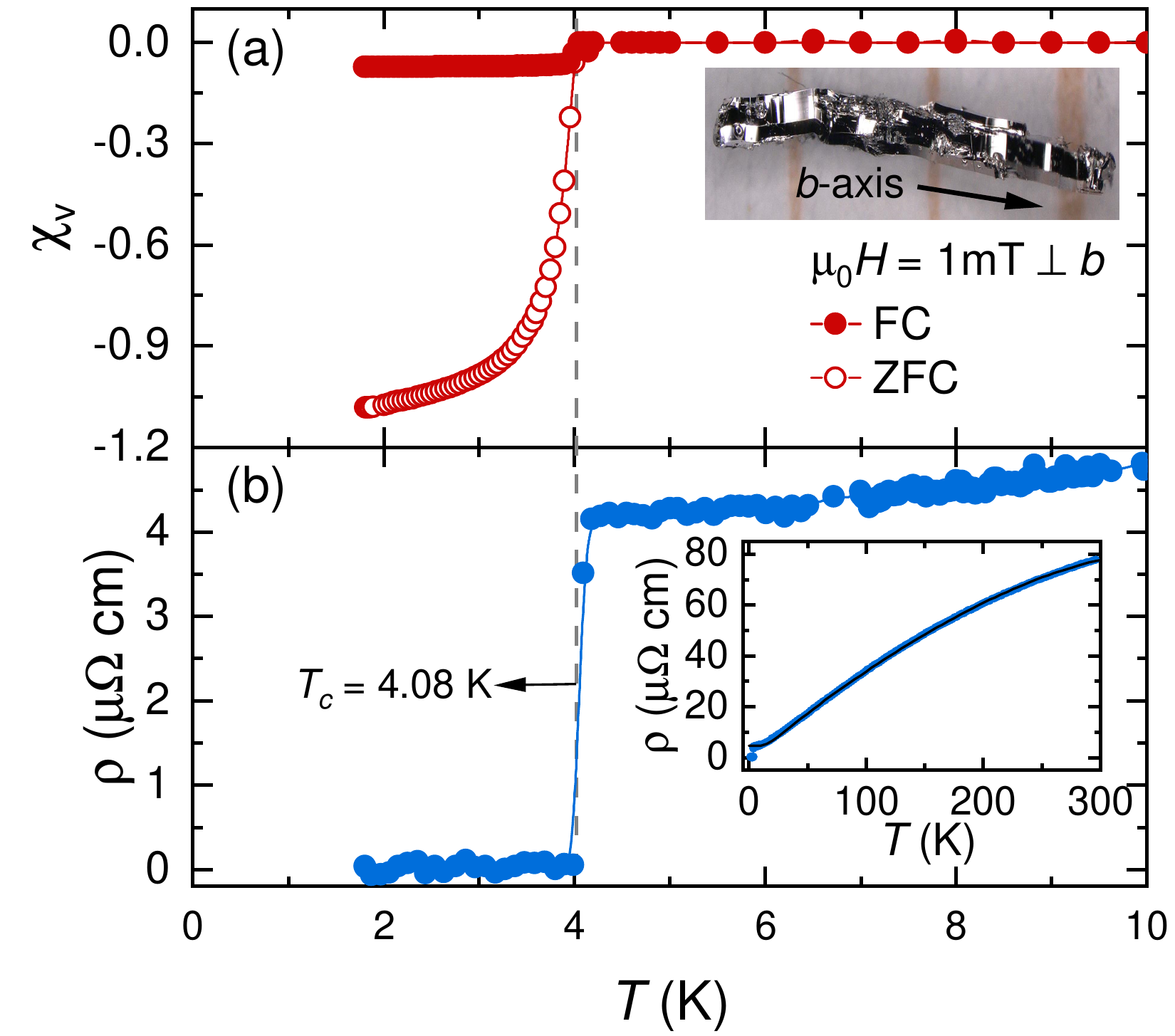}
	\caption{\label{fig:Tc}Temperature-dependent  magnetic susceptibility $\chi_\mathrm{V}(T)$ (a) and electrical resistivity $\rho(T)$ (b)  for a  NiBi$_3$ single crystal. 
		The dashed lines mark the $T_c$ value. 
		While $\rho(T)$ was measured in zero-field conditions, $\chi_\mathrm{V}(T)$ data 
		were collected in a magnetic field of 1\,mT, applied perpendicular to the $b$-axis. The susceptibility data 
		were corrected to account for the demagnetization factor.   
		The upper inset depicts a needle-like NiBi$_3$ single crystal, with the arrow indicating the $b$-axis, while the lower inset shows $\rho(T)$ up to 300\,K.
		In the latter case, the solid black line is a fit to the equation discussed in the text.}
\end{figure}
%


\section{Results and discussion\label{sec:results}}\enlargethispage{8pt}
\subsection{Crystal structure\label{ssec:XRD}}
The crystal structure and phase purity of NiBi$_3$ single crystals were checked via XRD at room temperature.
As shown in Fig.~\ref{fig:XRD}, we performed Rietveld refinements of the XRD pattern by using the \texttt{Fullprof} suite~\cite{Carvajal1993}. 
Consistent with previous results, we confirm that NiBi$_3$ crystallizes in
a centrosymmetric orthorhombic structure with a space group $Pnma$ (No.~62)~\cite{Fujimori2000,Kumar2011,Silva2013,Liu2020,Pineiro2011,Herrmannsdorfer2011,Zhu2012}. 
In this structure, bismuth atoms form an octahedral array, while nickel atoms form part of linear chains~\cite{Pineiro2011}.
The crystal structure of a unit cell is shown in the inset of Fig.~\ref{fig:XRD}(b). 
The refined lattice parameters, $a$ = 8.8799(1)\,\AA{}, $b$ = 4.09831(6)\,\AA{}, and $c$= 11.4853(1)\,\AA{}, 
are in good agreement with the results reported in the literature~\cite{Fujimori2000,Kumar2011,Silva2013,Liu2020,Pineiro2011,Herrmannsdorfer2011,Zhu2012}. 
According to the refinements in Fig.~\ref{fig:XRD}, a tiny amount of elemental bismuth ($\sim$1\%) was also identified, here attributed to the remanent 
Bi-flux on the surface of single crystals. Considering its small amount and the fact that Bi is not superconducting in the studied temperature range~\cite{Prakash2017}, 
the bismuth presence has negligible effects on the superconducting properties of NiBi$_3$.  
We also performed XRD measurement on a NiBi$_3$ single crystal. As shown in Fig.~\ref{fig:XRD}(b),
only the (004) reflection [the strongest among the (00$l$)-reflections]
was detected, confirming the single-crystal nature of the NiBi$_3$ sample.

\subsection{Bulk superconductivity\label{ssec:bulk_SC}}
\begin{figure}[!htp]
	\centering
	\includegraphics[width=0.48\textwidth,angle= 0]{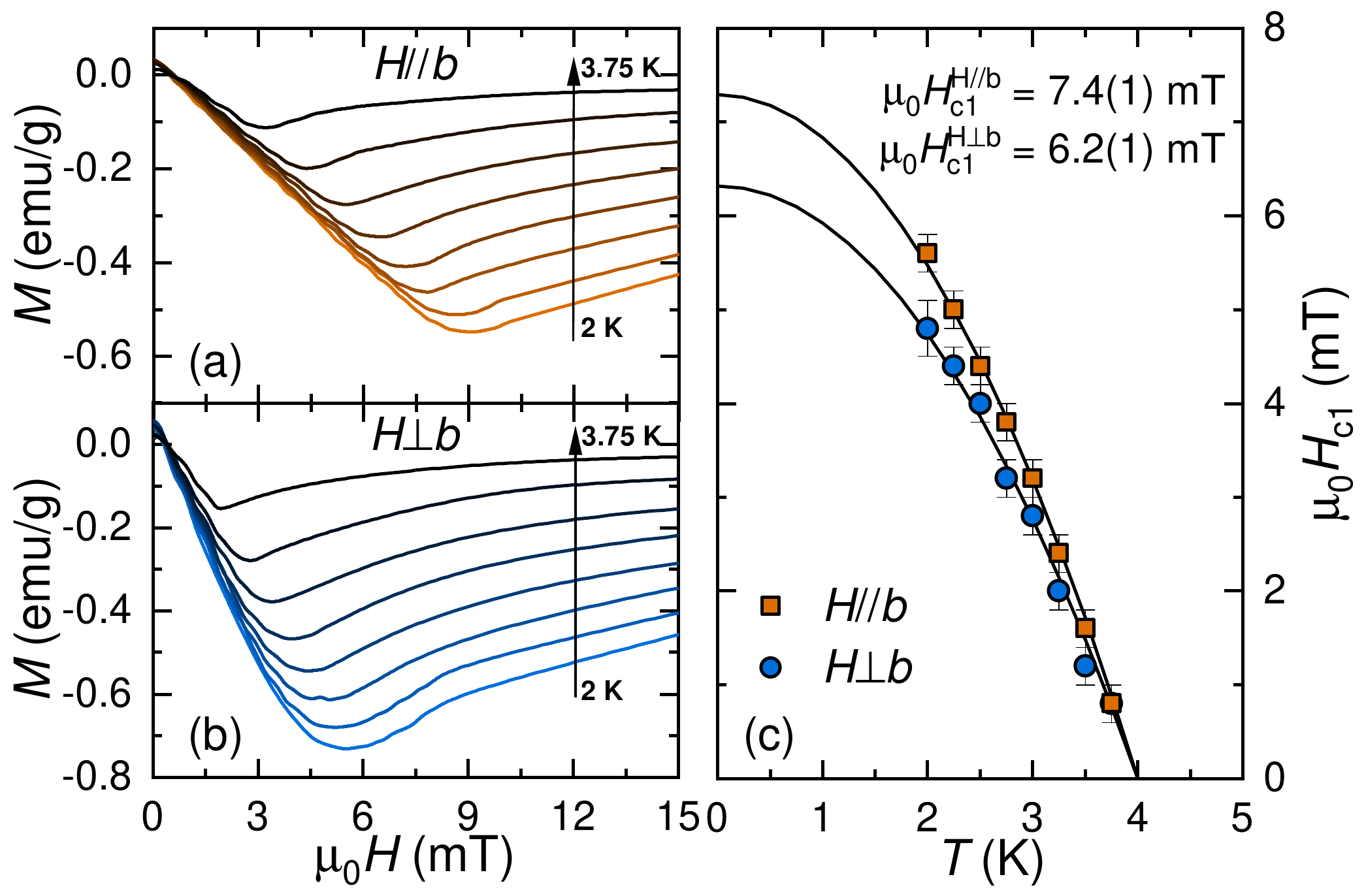}
	\caption{\label{fig:Hc1}Field-dependent magnetization curves
		collected at various temperatures after cooling the NiBi$_3$ single crystal in zero field for (a) $H \parallel b$ and (b) $H \perp b$.
		(c) The respective lower critical fields $H_\mathrm{c1}$ vs.\ temperature. Solid lines are fits to $\mu_{0}H_\mathrm{c1}(T) =\mu_{0}H_\mathrm{c1}(0)[1-(T/T_{c})^2]$.
		For each temperature, $H_\mathrm{c1}$ was determined as the 
		value where $M(H)$ starts deviating from linearity. The lower critical fields were corrected by a demagnetization factor $N$, 
			which is 0.5 for $H \perp b$. Since the NiBi$_3$ crystal is needle-like, no demagnetization factor is required for $H \parallel b$ (i.e., $N$ = 0).}
\end{figure}
The bulk superconductivity of NiBi$_3$ single crystals was
first characterized by magnetic-susceptibility measurements, using both 
field-cooling (FC) and zero-field-cooling (ZFC) protocols
in an applied field of 1\,mT. As shown in Fig.~\ref{fig:Tc}(a), 
a clear diamagnetic response
appears below the superconducting transition at 
$T_c$ = 4.05\,K. Similar results were obtained when applying the magnetic field parallel to the $b$-axis. 
After accounting for the demagnetization factor $N$, which is estimated to be 0.5 for $H \perp b$ and 0 for $H \parallel b$~\cite{Prozorov2018},  the superconducting 
shielding fraction of NiBi$_3$ was found to be close to 100\%, indicative of bulk SC.

The temperature-dependent electrical resistivity $\rho(T)$ of NiBi$_3$ single crystals was measured from 2\,K up to room temperature. It reveals 
a metallic behavior, without any anomalies associated with phase transitions in the normal state [see inset in Fig.~\ref{fig:Tc}(b)].
The electrical resistivity in the low-temperature region is plotted in the main panel of Fig.~\ref{fig:Tc}(b), where the superconducting
transition (with $T_c^\mathrm{onset} = 4.16$\,K, $T_c^\mathrm{mid} = 4.08$\,K and $T_c^\mathrm{zero} = 4.0$\,K) is clearly seen. 
As shown by the dashed-line in Fig.~\ref{fig:Tc}, the $T^\mathrm{mid}_c$ value is consistent with the onset of the superconducting transition in the magnetic susceptibility. Therefore, 
the $T^\mathrm{mid}_c$ values were used to determine the upper critical field of NiBi$_3$.   
The $\rho(T)$ curve can be described by the Bloch-Gr\"{u}neisen-Mott (BGM) formula 
$\rho(T) = \rho_0 + 4A \left(\frac{T}{\Theta_\mathrm{D}^\mathrm{R}}\right)^5\int_0^{\frac{\Theta_\mathrm{D}^\mathrm{R}}{T}}\!\!\frac{z^2\mathrm{d}z}{(e^z-1)(1-e^{-z})} - \alpha T^3$~\cite{Bloch1930,Blatt1968}.
Here, the first term $\rho_0$ is the residual resistivity due to the 
scattering of conduction electrons on the static defects of the crystal 
lattice, while the second term describes the electron-phonon scattering,  
with $\Theta_\mathrm{D}^\mathrm{R}$ being the characteristic (Debye) 
temperature and $A$ a coupling constant. The third term represents a 
contribution due to $s$-$d$ interband scattering, $\alpha$ being the 
Mott coefficient~\cite{Mott1958, Mott1964}.
The fit in Fig.~\ref{fig:Tc} (solid black line) results in 
$\rho_0 = 4.5(2)$\,\textmu$\Omega$cm, $A = 32(1)$\,\textmu$\Omega$cm, 
$\Theta_\mathrm{D}^\mathrm{R} = 100(3)$\,K, and 
$\alpha = 8.0(2)$\,$\times$10$^{-7}$\,\textmu$\Omega$cmK$^{-3}$. 
The derived $\Theta_\mathrm{D}^\mathrm{R}$ is comparable with the value determined from low-$T$ specific heat~\cite{Fujimori2000}.  
The fairly large residual resistivity ratio (RRR), i.e., $\rho$(300 K)/$\rho_0$ $\sim$ 18,
and the sharp superconducting transition ($\Delta$$T_c$ = 0.15\,K) both indicate a good sample quality.
\subsection{Lower and upper critical fields\label{ssec:critical_field}}

To determine the lower critical field $H_\mathrm{c1}$, the field-dependent magnetization $M(H)$ of a NiBi$_3$ single crystal was measured at 
various temperatures up to 3.75\,K. Figure~\ref{fig:Hc1}(a)-(b) plots the $M(H)$ for $H \parallel b$ and $H \perp b$, respectively.  
The estimated $H_\mathrm{c1}$ values at different temperatures 
(accounting for a demagnetization factor) are summarized in Fig.~\ref{fig:Hc1}(c).
The solid lines are fits to $\mu_{0}H_\mathrm{c1}(T) =\mu_{0}H_\mathrm{c1}(0)[1-(T/T_{c})^2]$ 
and yield the lower critical fields $\mu_{0}H_\mathrm{c1}(0)$ = 7.4(1) and 6.2(1)\,mT for $H \parallel b$ and $H \perp b$, respectively.
The zero-temperature $H_\mathrm{c1}^{H \parallel b}$/$H_\mathrm{c1}^{H \perp b}$ ratio ($\sim$1.2) suggests a weakly anisotropic superconductivity in the NiBi$_3$ single crystal.

\begin{figure}[!htp]
	\centering
	\includegraphics[width=0.49\textwidth,angle= 0]{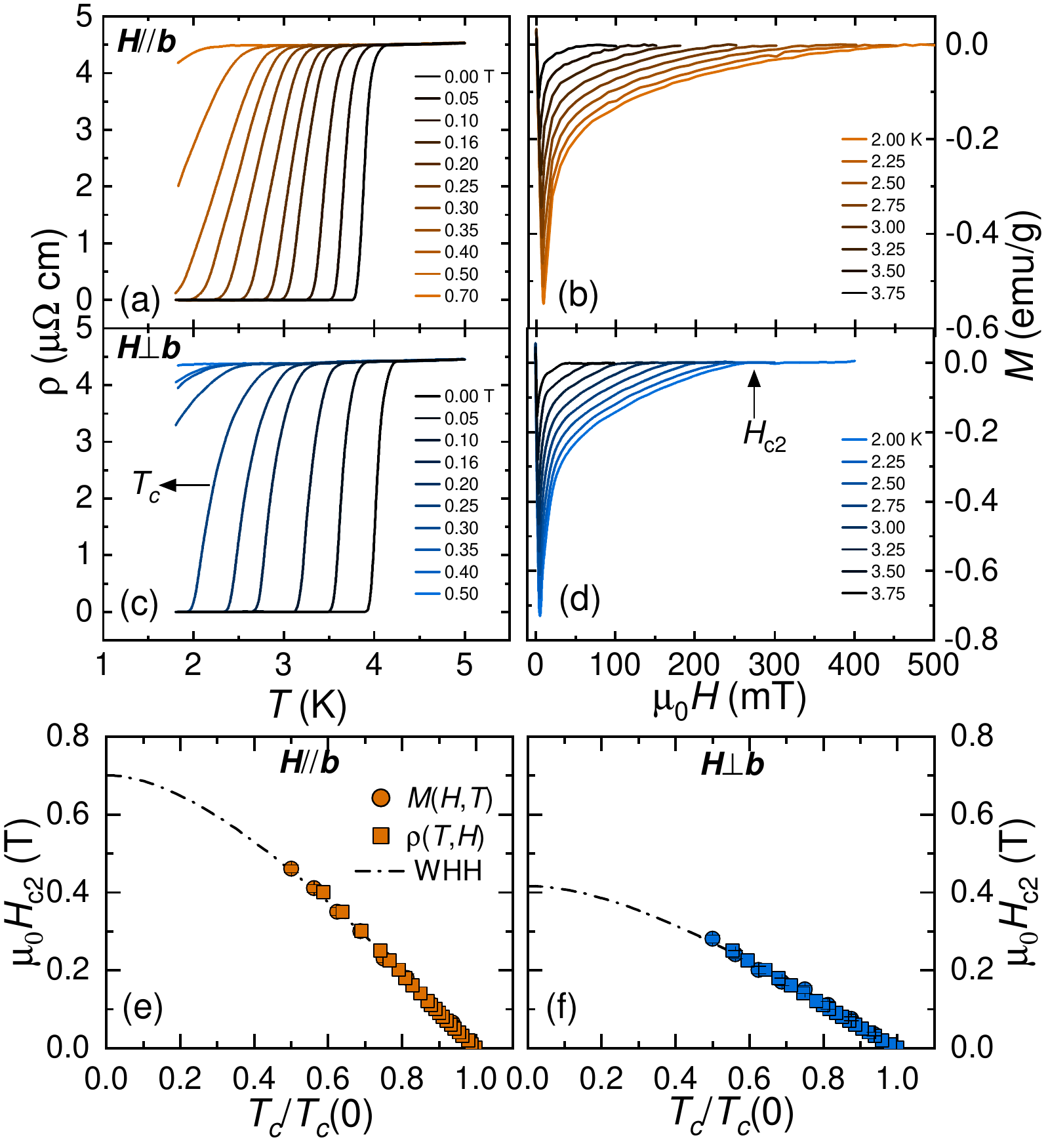}
	\caption{\label{fig:Hc2_raw}Temperature-dependent electrical resistivity $\rho(T,H)$ for various applied magnetic fields (a) 
		and field-dependent magnetization curves $M(H,T)$, collected at various temperatures below $T_c$ (b), for the $H \parallel b$ case.
	The analogue results for $H \perp b$ are shown in panels (c) and (d).
        In $\rho(T,H)$, $T_c$ was chosen as the mid-point of the superconducting transition (see details in Fig.~\ref{fig:Tc}). 
        While in $M(H,T)$,  $H_\mathrm{c2}$ was determined as the field where the diamagnetic 
		response vanishes (here indicated by an arrow).
Upper critical field $H_\mathrm{c2}$ vs.\ 
the reduced transition temperature $T_c/T_c(0)$ for $H \parallel b$ (e) and $H \perp b$ (f), respectively. 
The $T_c$ and $H_\mathrm{c2}$ values were determined from 
the measurements shown in panels (a)-(d). 
Circles refer to magnetization, while squares refer to 
resistivity measurements. The dash-dotted lines represent fits to the WHH model.	
}
\end{figure}

We also performed temperature-dependent electrical-resistivity measurements $\rho(T,H)$ 
at various applied magnetic fields, as well as studied the field-dependent 
magnetization $M(H,T)$ at various temperatures by applying the magnetic field either parallel or perpendicular to the $b$-axis of the NiBi$_3$ single crystal.
As shown in Fig.~\ref{fig:Hc2_raw}(a), upon increasing the magnetic 
field, the superconducting transition in $\rho(T)$ 
shifts to lower temperatures. 
Similarly, in the $M(H)$ data, the diamagnetic signal vanishes 
once the applied magnetic field exceeds the upper critical field 
$H_\mathrm{c2}$. For $H \parallel b$, no zero resistivity is observed down to $\sim2$\,K for $\mu_0H >$ 0.35\,T. For $H \perp b$, the zero resistivity already vanishes for $\mu_0H >$ 0.25\,T, implying that, in NiBi$_3$, the $H_\mathrm{c2}$ value along the $b$-axis is larger than that perpendicular to it.

Figure~\ref{fig:Hc2_raw}(e) and (f) summarizes the upper critical fields $H_\mathrm{c2}$ 
vs.\ the reduced superconducting transition temperature $T_c/T_c(0)$ for both $H \parallel b$ and $H \perp b$, as identified 
from the $\rho(T,H)$ and $M(H,T)$ data. The temperature evolution of the upper critical field  $H_\mathrm{c2}(T)$ is well described by the 
Werthamer-Helfand-Hohenberg (WHH) model~\cite{Werthamer1966}.
As shown by the dash-dotted lines in Fig.~\ref{fig:Hc2_raw}(e)-(f), the WHH model agrees remarkably well with the experimental data and provides $\mu_0H_\mathrm{c2}^{H \parallel b}(0)$ = 0.70(2)\,T and $\mu_0H_\mathrm{c2}^{H \perp b}(0)$ = 0.41(1), respectively. These values are comparable to previous results~\cite{Fujimori2000,Silva2013,Zhu2012} and both are much smaller than the Pauli-limiting field (i.e., 1.86$T_c$ $\sim$ 7.4\,T). Consequently, the orbital pair-breaking mechanism seems to be dominant in NiBi$_3$. Similarly to the lower critical fields, also the $H_\mathrm{c2}^{H \parallel b}$/$H_\mathrm{c2}^{H \perp b}$ ratio ($\sim$ 1.7) confirms the weakly
anisotropic SC in NiBi$_3$.

In the Ginzburg-Landau (GL) theory of superconductivity, the magnetic penetration 
depth $\lambda$ is related to the coherence length $\xi$, and the lower 
critical field via $\mu_{0}H_\mathrm{c1} = (\Phi_0 /4 \pi \lambda^2)$ln$(\kappa)$, 
where $\Phi_0 = 2.07 \times 10^{\tcb{3}}$\,T~nm$^{2}$ is the quantum of 
magnetic flux, $\kappa$ = $\lambda$/$\xi$ is the GL parameter~\cite{Tinkham1996}. 
By using $\mu_{0}H_\mathrm{c1}$ and $\xi$ values [calculated from $\mu_{0}H_\mathrm{c2}(0)$ = $\Phi_0$/2$\pi\xi(0)^2$], the resulting $\lambda_\mathrm{GL}(0) = 229(2)$
and 238(2)\,nm for $H \parallel b$ and  $H \perp b$ are both
compatible with the experimental value determined from $\mu$SR 
data. All the superconducting parameters are summarized in Table~\ref{tab:parameter}.
%
%
%
%
%
%
%
%

\subsection{TF-$\mu$SR and fully-gapped superconductivity\label{ssec:TF-muSR}}

\begin{figure}[!thp]
	\centering
	\includegraphics[width=0.48\textwidth,angle= 0]{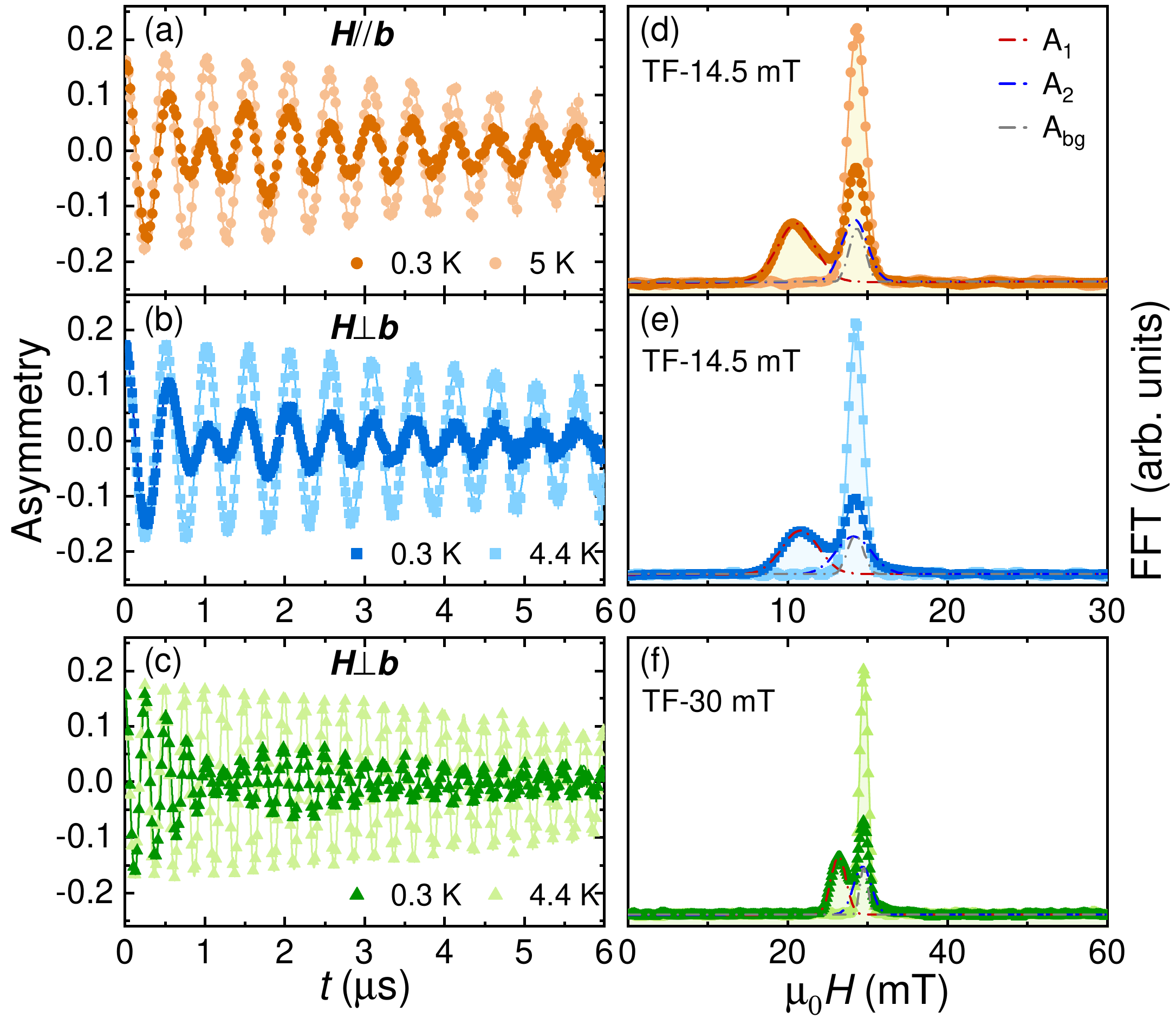}
	\caption{\label{fig:TF-muSR}TF-$\mu$SR spectra in the normal- and superconducting 
		states of NiBi$_3$ single crystals in an applied magnetic field of 14.5\,mT for $H \parallel b$ (a) and $H \perp b$ (b), respectively.
		The fast Fourier transforms of the TF-$\mu$SR spectra in panels (a) and (b) are shown in panels (d) and (e), respectively.  
		For $H \perp b$-axis, the TF-$\mu$SR and FFT spectra for an applied magnetic field of 30\,mT are shown in panels (c) and (f), respectively.  
		Solid lines are fits to Eq.~\eqref{eq:TF_muSR} using two oscillations, here also shown separately as
			dash-dotted lines in (c)-(f), together with a background contribution. Note the clear field-distribution broadening due to the onset of FLL below $T_c$.}
\end{figure}
%
%

To investigate the single-crystal anisotropy of superconducting pairing in NiBi$_3$, we carried 
out systematic transverse-field (TF-) $\mu$SR 
measurements, where we applied the field along different orientations. 
Representative TF-$\mu$SR spectra collected in a field of 14.5\,mT in the superconducting- and 
normal states of NiBi$_3$ are shown in Fig.~\ref{fig:TF-muSR}(a) for $H \parallel b$-axis and in Fig.~\ref{fig:TF-muSR}(b) for $H \perp b$-axis. 
In the latter case, the TF-$\mu$SR spectra were also collected in a field of 30\,mT [see Fig.~\ref{fig:TF-muSR}(c)].
In case of a type-II superconductor 
(see, e.g., the 0.3-K data in Fig.~\ref{fig:TF-muSR}), the development of a 
flux-line lattice (FLL) causes an inhomogeneous field distribution 
and, thus, it gives rise to an additional damping in the 
TF-$\mu$SR spectra~\cite{Yaouanc2011}.  
These are generally modeled using~\cite{Maisuradze2009}:
\begin{equation}
	\label{eq:TF_muSR}
	A_\mathrm{TF}(t) = \sum\limits_{i=1}^n A_i \cos(\gamma_{\mu} B_i t + \phi) e^{- \sigma_i^2 t^2/2} +
	A_\mathrm{bg} \cos(\gamma_{\mu} B_\mathrm{bg} t + \phi).
\end{equation}
Here $A_{i}$ ($\sim$85\%), $A_\mathrm{bg}$($\sim$15\%) and $B_{i}$, $B_\mathrm{bg}$ 
are the initial asymmetries and local fields sensed by the implanted muons in the 
sample and sample holder (i.e., Cu) or in the residual bismuth, $\gamma_{\mu}$/2$\pi$ = 135.53\,MHz/T 
is the muon gyromagnetic ratio, $\phi$ is a shared initial phase, and $\sigma_{i}$ is the Gaussian relaxation rate of the $i$th component. 
Note that, in the studied temperature range, the effects of the residual bismuth are negligible due to its non-magnetic and non-superconducting nature. 
We find that two oscillations (i.e., $n = 2$) are required
to describe the TF-$\mu$SR spectra for both $H \parallel b$- and $H \perp b$-axis
(see solid lines in Fig.~\ref{fig:TF-muSR}).
Indeed, as shown in Fig.~\ref{fig:TF-muSR}(a)-(c), oscillations with
two distinct frequencies can be clearly identified  
and, generally, the model with two oscillations provides a better fit 
than that with a single oscillation. For example, in
Fig.~\ref{fig:TF-muSR}(c), the two-oscillation model yields a
goodness-of-fit parameter $\chi_\mathrm{r}^2$ $\sim$ 1.4, twice 
smaller than the single-oscillation fit ($\chi_\mathrm{r}^2$ $\sim$ 3.2).
The fast Fourier transforms (FFT) of the TF-$\mu$SR spectra at 0.3\,K
are shown by dash-dotted lines in Fig.~\ref{fig:TF-muSR}(d)-(f), which
illustrate the two components ($A_1$ and $A_2$) and the background signal
($A_\mathrm{bg}$).
\tcr{The two-peak FFT in NiBi$_{3}$ might be related to two different
muon-stopping sites, a feature to be confirmed by future DFT calculations.}

\begin{figure}[!thp]
	\centering
	\includegraphics[width=0.48\textwidth,angle= 0]{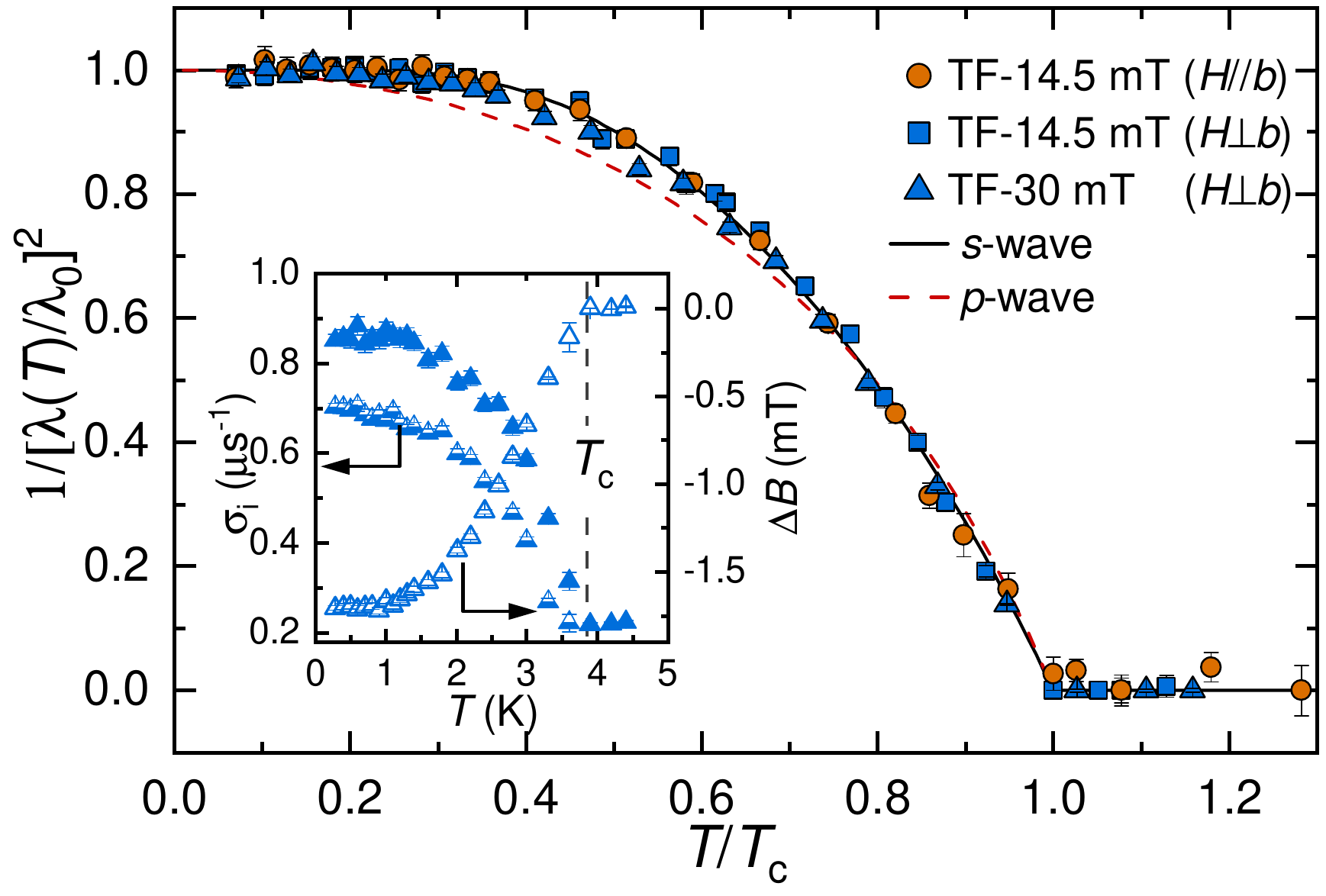}
	\caption{\label{fig:lambda}
		Superfluid density vs.\ reduced temperature $T/T_c$ for NiBi$_3$ single crystals. Inset shows the
		temperature-dependent muon-spin relaxation rate $\sigma_i(T)$ (left axis) and the diamagnetic shift $\Delta$$B(T)$ (right-axis) for the TF-30\,mT-$\mu$SR spectra. 
		The solid line in the main panel represents a fit to the fully gapped $s$-wave model with a single energy gap, while the dashed line is a fit using the $p$-wave model to the $H \parallel b$-data.}
\end{figure}

The derived muon-spin relaxation rates $\sigma_i$ are small and 
temperature-independent in the normal state, but below $T_c$ they start to increase due to the onset of FLL and the increased superfluid 
density.
At the same time, a diamagnetic field shift, $\Delta B (T) = \langle B \rangle - B_\mathrm{appl}$, appears below $T_c$,  
with $\langle B \rangle = (A_1\,B_1 + A_2\,B_2)/A_\mathrm{tot}$, where 
$A_\mathrm{tot} = A_1 + A_2$ and $B_\mathrm{appl}$ is the applied external field (see, e.g., the inset of Fig.~\ref{fig:lambda}).
The effective Gaussian relaxation rate can be calculated
from $\sigma_\mathrm{eff}^2/\gamma_\mu^2 = \sum_{i=1}^2 A_i [\sigma_i^2/\gamma_{\mu}^2 - \left(B_i - \langle B \rangle\right)^2]/A_\mathrm{tot}$~\cite{Maisuradze2009}.
Then, the superconducting Gaussian relaxation rate $\sigma_\mathrm{sc}$, can be extracted by subtracting 
the nuclear contribution according to $\sigma_\mathrm{sc} = \sqrt{\sigma_\mathrm{eff}^{2} - \sigma^{2}_\mathrm{n}}$. 
Here, $\sigma_\mathrm{n}$ is the nuclear relaxation rate, almost constant in the covered 
temperature range and relatively small in NiBi$_3$, as confirmed also by the ZF-$\mu$SR data (see Fig.~\ref{fig:ZF-muSR}). 
Since the applied TF fields (14.5 and 30\,mT) are not really small compared to the modest upper critical fields of NiBi$_3$ [see Fig.~\ref{fig:Hc2_raw}(e) and (f)],
to calculate the magnetic penetration depth $\lambda$ from $\sigma_\mathrm{sc}$ we had to consider the overlap of the vortex cores. 
Consequently, in our case, $\lambda$ was calculated by means
of $\sigma_\mathrm{sc} = 0.172 \frac{\gamma_{\mu} \Phi_0}{2\pi}(1-h)[1+1.21(1-\sqrt{h})^3]\lambda^{-2}$, where $h = B_\mathrm{appl}/B_\mathrm{c2}$~\cite{Barford1988,Brandt2003}.

Figure~\ref{fig:lambda} summarizes the temperature-dependent inverse 
square of the magnetic penetration depth [proportional to the 
superfluid density, i.e., $\lambda^{-2}(T) \propto \rho_\mathrm{sc}(T)$] for both $H \parallel b$ and $H \perp b$. For $H \perp b$, the data collected in a field of 30\,mT are also presented. 
The $\rho_\mathrm{sc}(T)$ was analyzed by applying different models, generally described by:
\begin{equation}
	\label{eq:rhos}
	\rho_\mathrm{sc}(T) = 1 + 2\, \Bigg{\langle} \int^{\infty}_{\Delta_\mathrm{k}} \frac{E}{\sqrt{E^2-\Delta_\mathrm{k}^2}} \frac{\partial f}{\partial E} \mathrm{d}E \Bigg{\rangle}_\mathrm{FS}. 
\end{equation}
Here, $f = (1+e^{E/k_\mathrm{B}T})^{-1}$ is the Fermi function~\cite{Tinkham1996}; 
$\Delta_\mathrm{k}(T) = \Delta(T) \delta_\mathrm{k}$ is an angle-dependent
gap function, where $\Delta$ is the maximum gap value and $\delta_\mathrm{k}$ is the 
angular dependence of the gap, equal to 1 and $\sin\theta$ 
for an $s$- and $p$-wave model, respectively, with $\theta$ being the azimuthal angles.
The temperature dependence of the gap is assumed to follow $\Delta(T) = \Delta_0 \mathrm{tanh} \{1.82[1.018(T_\mathrm{c}/T-1)]^{0.51} \}$~\cite{Tinkham1996,Carrington2003}, where $\Delta_0$ is the gap value at 0\,K.
As can be clearly seen in Fig.~\ref{fig:lambda} (see dashed line), the $p$-wave model exhibits a significant deviation from the experimental data. 
While the temperature-invariant superfluid density below $T_c$/3 strongly suggests the absence of low-energy excitations and, hence, a fully-gapped
superconducting state in NiBi$_3$. Consequently, the $\rho_\mathrm{sc}(T)$ is more consistent with the $s$-wave model.  
As shown by the solid line in Fig.~\ref{fig:lambda}, in both cases (i.e., $H \parallel b$- and $H \perp b$-axis), the $s$-wave model describes $\rho_\mathrm{sc}(T)$ very well across the entire
temperature range and yields a zero-temperature gap $\Delta_0$ = 2.10(5)\,$k_\mathrm{B}$$T_c$. 
Such isotropic gap value along different crystal directions is consistent with previous Andreev-reflection spectroscopy results~\cite{Zhao2018}.
For $H \parallel b$, the estimated $\lambda_0^\mathrm{ac}$ is 223(2)\,nm,  while for $H \perp b$, we find $\lambda_0^\mathrm{b}$ = 216(2) and 210(2)\,nm for TF-14.5\,mT and TF-30\,mT, respectively. 
The 
small $\lambda_0^\mathrm{ac}$/$\lambda_0^\mathrm{b}$ ratio (here, $\sim$ 1.03) and the similar gap sizes along different crystal directions confirm once more the weakly anisotropic SC 
of NiBi$_3$ single crystals. 
\tcr{
Note that, although the single-oscillation model also gives a temperature-independent superfluid density below 1/3$T_c$ as the two-oscillation model, the estimated $\lambda_0$ 
values by using the former model are significantly different from the values estimated from the magnetization data.}
\subsection{ZF-$\mu$SR and preserved time-reversal symmetry\label{ssec:ZF-muSR}}

\begin{figure}[t]
	\centering
	\includegraphics[width=0.49\textwidth,angle= 0]{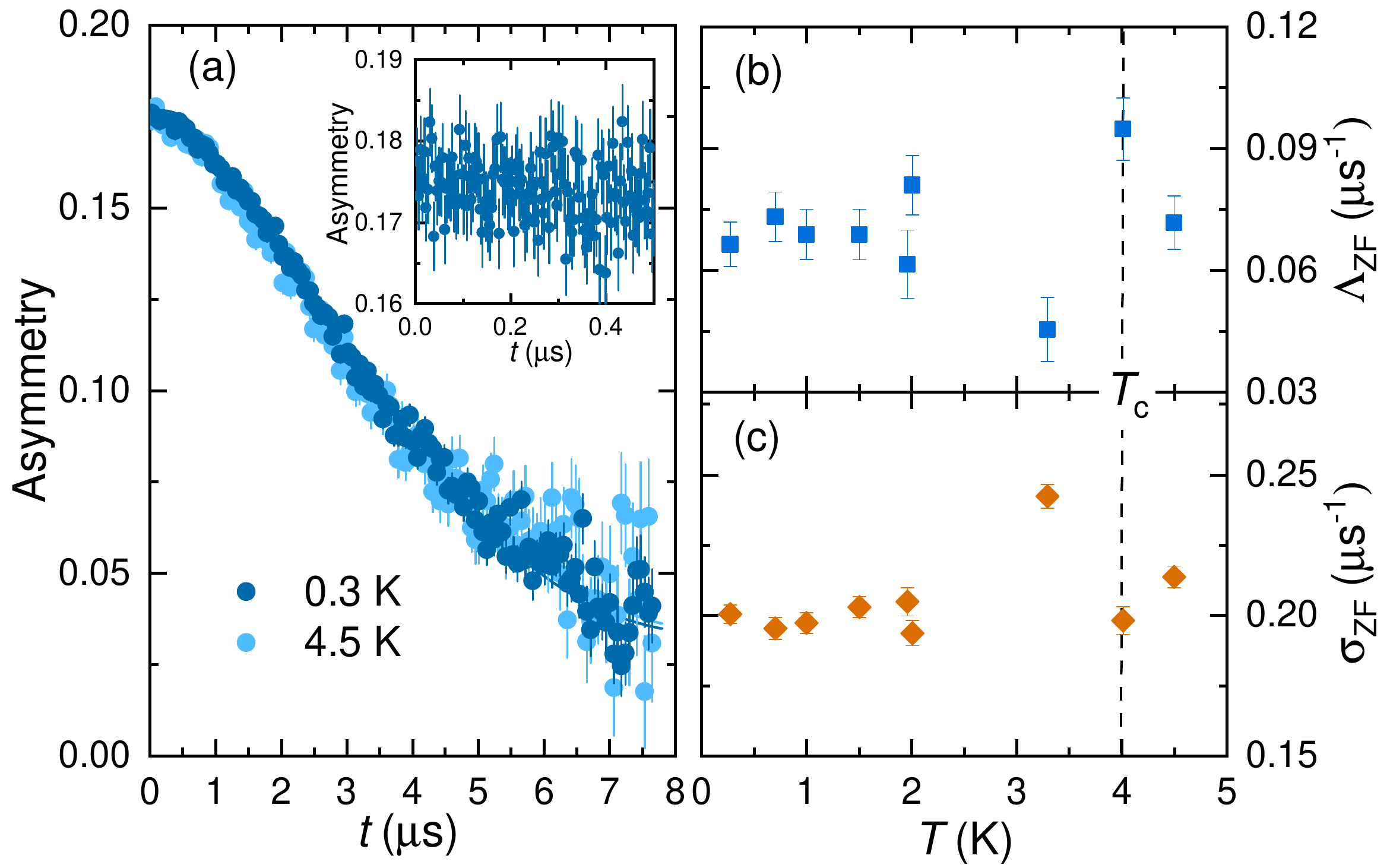}
	\caption{\label{fig:ZF-muSR}
		Representative ZF-$\mu$SR spectra of a NiBi$_3$ single crystal in the superconducting- (0.3\,K)
		and in the normal states (4.5\,K) (a). The inset shows an enlarged plot of 0.3\,K-ZF-$\mu$SR on a short time scale. 
		 Solid lines are fits to the equation described in the text, with the derived temperature-dependent Lorentzian- $\Lambda_\mathrm{ZF}(T)$ and Gaussian $\sigma_\mathrm{ZF}(T)$ relaxation rates being shown in panels (b) and (c), respectively.
        The dashed line at 4\,K marks the bulk $T_c$ and serves as a guide to the eyes. None of the reported fit parameters shows a 
        distinct enhancement below $T_c$, thus suggesting a preserved TRS.}
\end{figure}

To search for possible ferromagnetism and breaking of the time-reversal symmetry in a NiBi$_3$ single crystal, we performed zero-field (ZF-) $\mu$SR measurements covering both the normal- and superconducting states. 
As shown in Fig.~\ref{fig:ZF-muSR}, neither coherent oscillations nor fast decays could be identified in 
the spectra collected above- (4.5\,K) and below $T_c$ (0.3\,K), hence implying the lack of any magnetic order or fluctuations. 
As clearly demonstrated in the inset of Fig.~\ref{fig:ZF-muSR}(a), the ZF-$\mu$SR spectra are
 almost flat on a short time scale (i.e., $t$ < 0.5\,$\mu$s), confirming again the absence of fast oscillations which would be caused by a 
possible ferromagnetic ordering, as instead reported in previous work~\cite{Liu2020,Pineiro2011}.    
Thus, in the absence of external fields, the muon-spin relaxation in NiBi$_3$ is mainly due to the randomly 
oriented nuclear moments, which can be modeled by a Gaussian Kubo-Toyabe relaxation function
$G_\mathrm{KT} = [\frac{1}{3} + \frac{2}{3}(1 -\sigma_\mathrm{ZF}^{2}t^{2})\,\mathrm{e}^{-\sigma_\mathrm{ZF}^{2}t^{2}/2}]$
~\cite{Kubo1967,Yaouanc2011}.
Here, $\sigma_\mathrm{ZF}$ is the zero-field Gaussian relaxation rate. 
The ZF-$\mu$SR spectra were fitted by considering also an additional electronic contribution. 
The solid lines in Fig.~\ref{fig:ZF-muSR}(a) represent fits to $A_\mathrm{ZF}(t) = A_\mathrm{s} G_\mathrm{KT} \mathrm{e}^{-\Lambda_\mathrm{ZF} t} + A_\mathrm{bg}$, where $\Lambda_\mathrm{ZF}$ is the zero-field exponential muon-spin relaxation rate, $A_\mathrm{s}$ and $A_\mathrm{bg}$ are the same as in TF-$\mu$SR [see Eq.~\eqref{eq:TF_muSR}].
The derived $\Lambda_\mathrm{ZF}$ and  $\sigma_\mathrm{ZF}$ values as a function of temperature are shown in Fig.~\ref{fig:ZF-muSR}(b) and (c),  respectively.  
Neither $\Lambda_\mathrm{ZF}(T)$ nor $\sigma_\mathrm{ZF}(T)$ show a systematic enhancement below $T_c$.
Here, the jump of $\sigma_\mathrm{ZF}$ at 3.5\,K [Fig.~\ref{fig:ZF-muSR}(c)] is not related to a TRS-breaking effect, but to the correlated
decrease of $\Lambda_\mathrm{ZF}$ [Fig.~\ref{fig:ZF-muSR}(b)].
The lack of an additional relaxation below $T_c$ excludes a possible TRS-breaking effect in the superconducting state of the NiBi$_3$ single crystal, here also reflected in the practically overlapping datasets shown in Fig.~\ref{fig:ZF-muSR}(a). 
\section{\label{ssec:dis}Discussion}\enlargethispage{8pt}
%
\begin{figure}[ht]
	\centering
	\includegraphics[width = 0.48\textwidth]{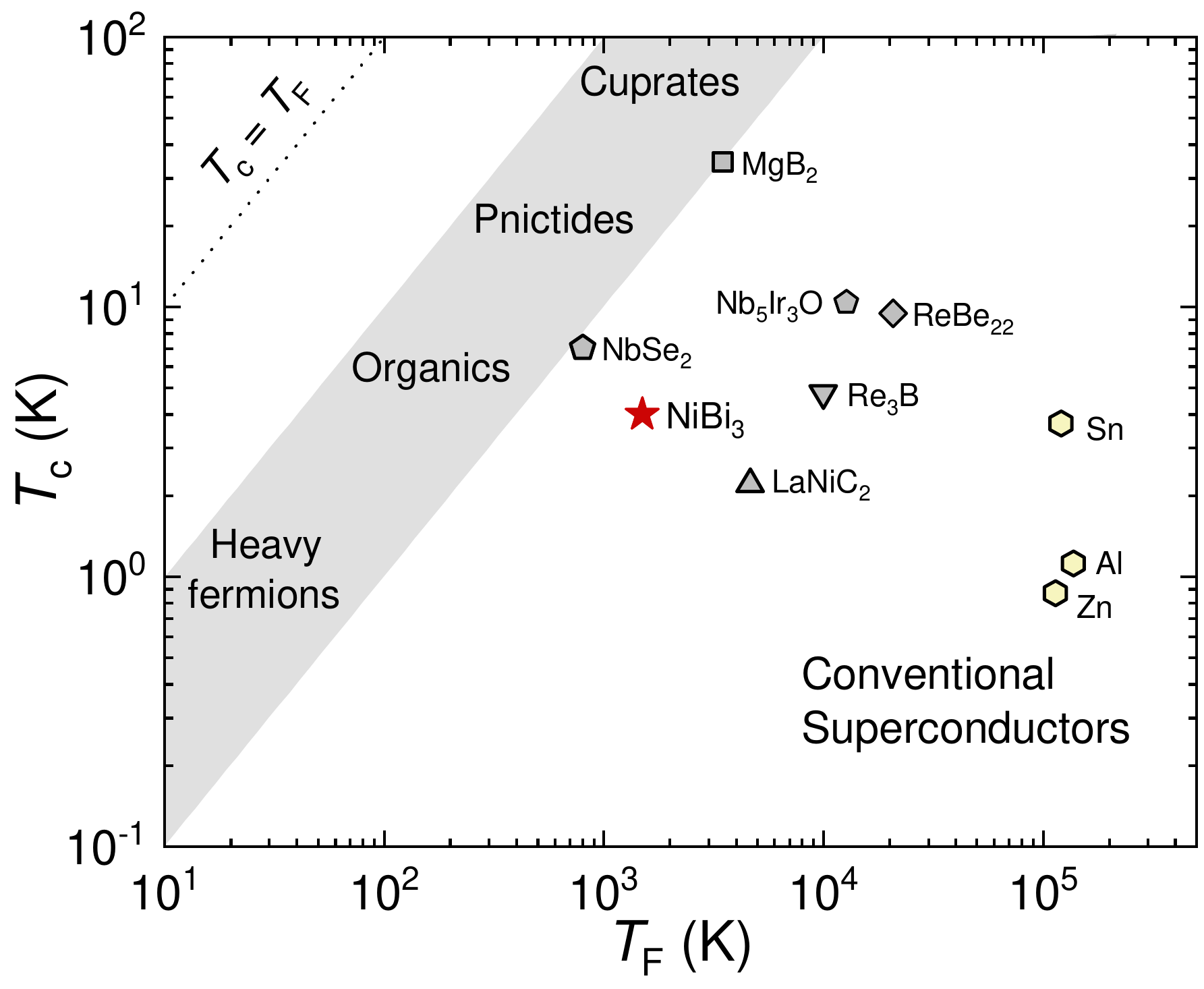}
	\caption{\label{fig:uemura}Uemura plot showing the superconducting transition $T_c$ against the effective Fermi temperature $T_\mathrm{F}$ for different kinds of superconductors. The grey region $1/100 < T_c/T_\mathrm{F}<1/10$ delimits the band of 
		unconventional superconductors, including heavy fermions, organics, fullerenes, 
		pnictides, and high-$T_c$ cuprates. A few selected superconductors (see Refs.~\onlinecite{Shang2019c,Uemura1991,Uemura2006,Shang2020NbIr,Shang2019b})
		are shown here with different symbols. The dotted line  
		corresponds to $T_c = T_\mathrm{F}$, where $T_\mathrm{F}$ is the 
		temperature associated with the Fermi energy, $E_\mathrm{F} = k_\mathrm{B} T_\mathrm{F}$.}
\end{figure}
%
First, we discuss the absence of ferromagnetism in NiBi$_3$. Previous works found evidence for either extrinsic or intrinsic ferromagnetism in NiBi$_3$ single crystals. 
In the extrinsic case, the amorphous Ni impurities in NiBi$_3$ lead to a clear drop in magnetization at temperatures close to their Curie temperature~\cite{Silva2013}.
In NiBi$_3$, such FM can even coexist with SC, as demonstrated by the ferromagnetic hysteresis loops observed in the magnetization data~\cite{Liu2020,Pineiro2011,Herrmannsdorfer2011}. 
Besides bulk crystals with Ni impurities, also submicrometer-sized NiBi$_3$ particles, or quasi-one-dimensional nanoscale strips of NiBi$_3$ undergo a ferromagnetic order~\cite{Herrmannsdorfer2011}. 
In addition, below 150\,K, electron-spin-resonance data reveal  ferromagnetic-like fluctuations on the surface of NiBi$_3$ crystals~\cite{Zhu2012}. 
The above evidence suggests that NiBi$_3$ is a possible ferromagnetic superconductor, which might enable spin-triplet SC pairing, analogous to that found in other ferromagnetic 
superconductors~\cite{Saxena2000,Aoki2001,Huy2007}.
Here, we applied the ZF-$\mu$SR technique to detect a possible ferromagnetic order in NiBi$_3$ single crystals. 
Due to their large magnetic moment,  
muons represent one of the most sensitive magnetic probes, able to sense very low internal fields ($\sim$$10^{-2}$\,mT), and thus, to detect local magnetic fields of either nuclear or electronic origin~\cite{Yaouanc2011,Amato1997}. 
In contrast to previous studies, our ZF-$\mu$SR results do not evidence any ferromagnetic order or fluctuations in NiBi$_3$, thus implying that the previously reported ferromagnetism is most likely of extrinsic nature.
In addition, our ZF-$\mu$SR results indicate that time-reversal symetry is preserved in the superconducting state of NiBi$_3$. This conclusion is further supported by recent high-resolution surface magneto-optic Kerr effect measurements in NiBi$_3$ single crystals, which also do not find any traces of a spontaneous Kerr signal in the superconducting state~\cite{Wang2022}.

Second, we discuss the weakly anisotropic superconductivity in NiBi$_3$.  
Although NiBi$_3$ single crystals exhibit a quasi-one dimensional needle-like shape [see inset in Fig.~\ref{fig:Tc}(a)], their crystal structure is definitely three dimensional, with $a$ =8.8799\,\AA{}, $b$ = 4.09831\,\AA{}, and $c$= 11.4853\,\AA{} (see Fig.~\ref{fig:XRD}). Such an intrinsically three-dimensional structure could naturally explain the almost isotropic- (or weakly anisotropic) electronic properties of NiBi$_3$. 
As confirmed by our study, the lower critical field $H_\mathrm{c1}$ (Fig.~\ref{fig:Hc1}), the upper critical field $H_\mathrm{c2}$ (Fig.~\ref{fig:Hc2_raw}), the magnetic penetration depth $\lambda$, and the superconducting gap $\Delta_0$ (Fig.~\ref{fig:lambda}), all show comparable values for $H \parallel b$- and $H \perp b$-axis. These results clearly indicate a weakly anisotropic SC in NiBi$_3$ single crystals, as also
observed in NiBi$_3$ thin films~\cite{Wang2018}.  
Moreover, electrical-resistivity measurements reveal almost identical upper critical fields when the magnetic field is rotated in the $ac$-plane of a NiBi$_3$ single crystal~\cite{Fujimori2000}.
Future electronic band-structure calculations could provide further hints about the observed weak anisotropy.

Finally, let us compare the superconducting parameters of NiBi$_3$ with those of other superconductors. By using the SC parameters obtained from the measurements presented here, we calculate an effective Fermi temperature $T_\mathrm{F}$ $\sim$ 1.5 $\times$10$^3$\,K for NiBi$_3$ (here, we consider the $H \parallel b$ case). We recall that $T_\mathrm{F}$ is proportional to $n_\mathrm{s}^{2/3}$/$m^\star$, where $n_\mathrm{s}$ and $m^\star$ are the carrier density and the effective mass. The estimated $m^\star$ $\sim$ 8\,$m_e$, indicates a moderate degree of electronic correlations in NiBi$_3$.  
Different families of superconductors can be classified according to their $T_c$/$T_\mathrm{F}$ ratios into a so-called Uemura plot~\cite{Uemura1991}.
In such plot, many conventional superconductors, as e.g., Al, Sn, and Zn, are located at $T_c/T_\mathrm{F} < \sim$10$^{-4}$. 
Conversely, several types of unconventional superconductors, including heavy fermions, organics, high-$T_c$ iron pnictides, and cuprates, all lie in the 10$^{-2}$ $< T_c/T_\mathrm{F} <$ 10$^{-1}$ band. 
Between these two categories lie a few different types of superconductors, including multigap- and noncentrosymmetric superconductors, etc.~\cite{Shang2019c,Uemura1991,Uemura2006,Shang2020NbIr,Shang2019b}. 
Although located clearly far off the conventional band, our TF- and ZF-$\mu$SR results suggest that the NiBi$_3$ superconductivity is more consistent with a  conventional fully-gapped SC with preserved time-reversal symmetry. Such conclusion is also supported by the linear suppression of $T_c$ under applied pressure. For instance, the $T_c$ of NiBi$_3$ decreases from 4\,K to 3\,K when the pressure increases from 0 to 2.2\,GPa~\cite{Gati2019}. 

\begin{table}[!th]
	\centering
	\caption{Superconducting parameters of NiBi$_3$ single crystal, as 
			determined from electrical-resistivity, magnetization, and TF-14.5\,mT-$\mu$SR measurements. 
	    \label{tab:parameter}} 
	\begin{ruledtabular}
			\begin{tabular}{lccc}
			Property                                 &  Unit             & $H \parallel b$        & $H \perp b$      \\ \hline
			$T_c^\chi$                               & K                  & 4.05(5)                  &  4.05(5)          \\
			$T_c^{\mu\mathrm{SR}}$                   & K                  & 3.9(1)                   &  3.9(1)        \\
			$\mu_0H_{c1}$                            & mT                 & 7.4(1)                & 6.2(1)            \\
			$\mu_0H_{c2}$                            & T                  & 0.70(1)               & 0.41(1)          \\
			$\Delta_0$($s$-wave)                     & $k_\mathrm{B}T_c$                & 2.10(5)               & 2.10(5)       \\  
			$\Delta_0$($p$-wave)                     & $k_\mathrm{B}T_c$                & 3.00(5)               & 3.00(5)         \\  
			$\lambda_0$                              & nm                 & 223(2)                & 216(2)           \\
			$\xi(0)$                                 & nm                 & 21.7(2)               & 28.3(3)          \\
			$\lambda_\mathrm{GL}(0)$                 & nm                 & 229(2)                & 238(2)           \\
			$\kappa$                                 & --                 & 10.5(2)               & 8.4(2)          \\
		\end{tabular}	
	\end{ruledtabular}
\end{table}

\vspace{3pt}
\section{\label{ssec:Sum}Conclusion}\enlargethispage{8pt}
To summarize, we investigated the superconducting properties
of flux-grown NiBi$_3$ single crystals by means of 
electrical resistivity-, magnetization-, and $\mu$SR measurements. NiBi$_3$ was shown to exhibit bulk SC with $T_c$ $\sim$ 4.1 K. 
By applying magnetic fields along different crystal directions, we obtained the
lower critical field $H_\mathrm{c1}$, the upper critical field $H_\mathrm{c2}$, and the magnetic penetration depth $\lambda$ for both the $H \parallel b$-  and $H \perp b$ cases.  
Both $H_\mathrm{c2}^{H \parallel b}(0)$ and $H_\mathrm{c2}^{H \perp b}(0)$ values are much smaller than the Pauli-limit 
field, implying that the orbital pair-breaking mechanism is dominant in NiBi$_3$.  
Although the NiBi$_3$ single crystals show needle-like shapes, their superconducting properties are only weakly anisotropic, as reflected in
the small
ratios of $H_\mathrm{c1}^{H \parallel b}$/$H_\mathrm{c1}^{H \perp b}$, $H_\mathrm{c2}^{H \parallel b}$/$H_\mathrm{c2}^{H \perp b}$, and $\lambda_0^\mathrm{ac}$/$\lambda_0^\mathrm{b}$. The temperature dependence of the NiBi$_3$ superfluid density reveals a \emph{nodeless} SC, well described by an isotropic $s$-wave model. The relatively large gap value, here, $\Delta_0$ = 2.1 $k_\mathrm{B}T_c$, suggests strong electron-phonon interactions in NiBi$_3$. Further, the lack of spontaneous magnetic fields below $T_c$ indicates that the time-reversal symmetry is preserved in the NiBi$_3$ superconductor. The absence of a fast relaxation and/or oscillations in the ZF-$\mu$SR 
spectra excludes a magnetic order on the surface or in the bulk of NiBi$_3$ single crystals, thus implying that the previously observed ferromagnetism is most likely of extrinsic origin. 
Overall, our systematic results suggest that, in contrast to the unconventional superconductivity observed in Bi/Ni bilayers, NiBi$_3$ behaves as a conventional $s$-type superconductor.


\vspace{1pt}
\begin{acknowledgments}
This work was supported by the Natural Science Foundation of Shanghai 
(Grants No.\ 21ZR1420500 and 21JC\-140\-2300), Natural Science
Foundation of Chongqing (Grant No.\ 2022NSCQ-MSX1468), and the Schweizerische 
Nationalfonds zur F\"{o}r\-der\-ung der Wis\-sen\-schaft\-lichen For\-schung 
(SNF) (Grants No.\ 200021\_188706 and 206021\_139082). Y.X.\ acknowledges
support from the Shanghai Pujiang Program (Grant No.\ 21PJ1403100).
We acknowledge the allocation of beam time at the Swiss muon source 
(Dolly $\mu$SR spectrometer).
\end{acknowledgments}


\bibliography{NiBi3.bib}

\end{document}